\documentclass{styles/svproc}
\usepackage[ruled,vlined,linesnumbered,longend,resetcount]{algorithm2e}

\usepackage{amsmath}
\usepackage{mathtools}
\usepackage{tabularx}
\usepackage{amsfonts}
\usepackage{subcaption}
\usepackage[labelsep=period]{caption}
\usepackage{subcaption}
\usepackage{graphicx}
\usepackage{lipsum}
\usepackage{float}

\usepackage{url}

\begin{document}
\mainmatter              % start of a contribution
\title{Minimal Sleep Delay Driven Aggregation Tree Construction in IoT Sensor Networks}
\titlerunning{Minimal Sleep Delay Driven Aggregation Tree}  % abbreviated title (for running head)
\author{Van-Vi Vo\inst{1} \and Duc-Tai Le\inst{2} \and Hyunseung Choo\inst{1}}
\authorrunning{Van-Vi Vo et al.} % abbreviated author list (for running head)
\tocauthor{Van-Vi Vo, Duc-Tai Le, and Hyunseung Choo}
\index{Vo, V.}
\index{Le, D.}
\index{Choo, H.}
\institute{Department of Electrical and Computer Engineering, \\
\and
College of Computing and Informatics, \\
Sungkyunkwan University, Suwon, South Korea, \email{choo@skku.edu\\}}

\maketitle              % typeset the title of the contribution

\begin{abstract}
Data aggregation is a fundamental technique in wireless sensor networks (WSNs) in which sensory data collected by intermediate nodes is merged by in-network computation using maximum, average, or sum functions. Because sensors run on batteries, energy conservation is a critical issue. Duty cycle is a well-known energy-saving mechanism in WSNs, but it causes data aggregation latency to increase. As a result, the use of multichannel technology allows more sensor nodes to send data simultaneously, reducing data aggregation latency. We investigate the minimum latency aggregation scheduling problem in multi-channel duty-cycled IoT sensor networks in this paper. We propose a scheduling scheme that first constructs an aggregation tree based on sensor node sleep delay, then improves parallel transmissions by scheduling all eligible nodes in the constructed aggregation tree to enhance data aggregation. Based on extensive simulation experiments, our proposed approach lowers the aggregation delay by at most 61\% compared to a novel approach.

\end{abstract}

\section{Introduction}
Wireless sensor networks (WSNs) are becoming more prevalent in Internet of Things (IoT) applications. A WSN is composed of a number of sensor nodes that are typically deployed in inaccessible areas, such as undersea, underground, or within the human body, to collect and measure data, which is then collected by a sink or base station \cite{Abbasian Dehkordi}. Sensors are battery-powered and have limited computation capabilities. Data communication between intermediate nodes, on the other hand, necessitates a significant amount of energy. Data aggregation is an effective method for integrating data to address this issue. This technique reduces redundancy in sensory data. Furthermore, it has an effect on data aggregation delay.

While aggregating data, sensors in the network do not constantly send and receive data. Keeping the network active while not performing any tasks wastes a lot of energy and shortens the network's lifespan. To address this issue, a duty-cycling mechanism is used to convert energy when the sensor nodes are idle \cite{Wen Yean-Fu}. When sending or receiving data, sensors in the duty-cycling technique are active, and when idle, they are inactive. When sensors are not in use, they save energy. Because the inactive period is longer than the wake-up period, the technique conserves more energy, but the data aggregation delay increases.

Time and energy efficiency in data aggregation are major concerns in WSNs. Many studies have been conducted to investigate energy-efficient and time-efficient data aggregation problems in WSNs \cite{KhadirKumar}, \cite{Chen Quan}, \cite{Yu Bo}, \cite{Chen Quan 2}. The researchers of \cite{KhadirKumar},  \cite{Chen Quan}, \cite{Yu Bo} investigate how to improve sensor network energy while minimizing latency in data aggregation sensor networks. KhadirKumar, et al \cite{KhadirKumar} propose a data aggregation strategy that produced more efficient data aggregation, throughput, and lifetime results. In contrast to \cite{Chen Quan} which addresses the problem in battery-free sensor networks, the authors propose three data aggregation algorithms considered energy-collision aware that build an adaptable aggregation tree that takes into account residual and harvested energy in each sensor to reduce aggregation latency.

The minimum latency data aggregation (MLAS) problem in duty-cycled WSNs is first investigated in \cite{Yu Bo}, which has been shown to be NP-hard, and they use connected dominating set (CDS) to construct the routing structure. The authors of \cite{Chen Quan 2} propose a distributed algorithm using multichannel on an aggregation tree to reduce latency. The use of multiple channels helps to increase the number of concurrent transmissions at each time schedule. In this paper, we propose a time-efficient data aggregation scheduling algorithm for multichannel duty-cycled sensor networks. One study uses CDS to investigate reducing latency for data aggregation in multichannel duty-cycled wireless sensor networks \cite{Jiao Xianlong}. In this paper, we propose a new method for building the aggregation tree based on the active time slots of sensor nodes, as well as an improved scheduling scheme. We conduct intensive simulation experiments to examine the effectiveness of the proposed scheme with the reference. In terms of latency, the proposed algorithm outperforms the reference scheme.

The rest of the paper is structured as follows. The network model, assumptions, and problem formulation are presented in Section 2. Section 3 then goes over our proposed scheme. Section 4 compares the experimental results of the proposed scheme to the modern approach. Finally, in Section 5, we wrap up our work and plan for the future work.

\section{Preliminaries}
\subsection{Network model and assumptions}
Consider the undirected graph $G = (V,E)$, where $V$ be a set of sensor nodes including a sink node $s$ and $E$ is a set of communication links between nodes. Each sensor node has a uniform transmission range $d$ and is outfitted with an omnidirectional antenna for wireless communications. According \cite{Shi Yi}, we employ the interference model, in which sensor nodes, due to their half-duplex mode, cannot transmit and receive data at the same time. An interference occurs when a sensor $u$ sends data to a receiver $v$ while $v$ is within the interference range of another unintended concurrent transmission. The interference range of a sensor node is denoted by $d_I$, $d_I \geq d$. There is a set of $m$ given channels $F = {f_1,f_2,…,f_m}$ can be used to allocate to sensor nodes. The usage of different channels for sensor nodes increases interference-free concurrent transmissions. Consequently, the total aggregation delay is reduced. 

To save energy, we assume that sensor nodes use duty-cycled mechanisms like \cite{Chen Quan 3}. The sensor nodes in this mechanism switch between sleeping and active modes independently. Sensor node network lifetime is split into working periods with the same length $T$. Each working period is further divided into $W = {0,1,…,(T-1)}$ time slots. Each time slot is assumed to be a sufficient time for a successful data transmission. Every sensor node has its own sleeping and active time slots in a working period, and the sleeping and active patterns of sensors repeat in every working period throughout their entire lifetimes. The sensors are inactive during their sleeping time slots and awaken to receive data during their active time slots. Furthermore, they wake up at any time during the scheduled time slot to transmit data. In this paper, we assume that during a working period, each sensor node $u$ is active at random and independently at $\alpha$ time slots. The duty cycle is defined as the ratio of active slots number $\alpha$ to the length of the working period $T$, i.e., $\alpha/T$.

\subsection{Problem formulation}
In this paper, we study Minimum Latency Aggregation Scheduling (MLAS) problem in multi-channel duty-cycled IoT sensor networks. The data aggregation completes when sink node $s$ collects data from all sensor nodes in the network. In other words, all nodes in the network except the sink must be assigned channels and time slots when scheduling completes to transmit data with interference-free. Data aggregation delay is the number of time slots needed to aggregate data by the sink from all nodes in the network. The target of MLAS is to reduce data aggregation delay while also ensuring that all scheduled transmissions are interference-free.

\section{Proposed data aggregation scheduling algorithms}
The proposed data aggregation scheduling scheme is described in this section, it is divided into two phases: Aggregation tree construction phase which determines directional sender-receiver (child-parent) pairs for data transmissions; and data scheduling phase which determines the time for sensor nodes to transmit data without interference based on the constructed tree in the first phase by allocating channels and assigning time slots to sensors. We define a sleep delay metric is a time interval before a transmission between sender $u$ and receiver $v$ at active slots $\tau_u$ and $\tau_v$ in a working period length $T$ as follows:

\begin{equation}
    t(\tau_u, \tau_v, T) =
    \begin{cases}
      \tau_v - \tau_u, & \text{if } \tau_v > \tau_u \\
      \tau_v + T - \tau_u, & \text{otherwise.}
    \end{cases}
\end{equation}

The sleep delay measures the time for node $u$ to wait until receiver $v$ active so that $u$ can transmit data to $v$. Node $u$ transmits data to $v$ in the same working period when receiver $v$ is active at later active slot of $u$ and $u$ must wait for a new working period to send data to receiver $v$ if receiver $v$ is active before $u$. Intuitively, minimizing sleep delay between sender $u$ and receiver $v$ also reduces sleep delay waiting for transmissions of other nodes. As a result, the total latency will be reduced. Consequently, we define minimal sleep delay between sender $u$ and receiver $v$ as follows:

\begin{equation}
    d(u,v, T) = \underset{\tau_u \in A(u)}{\mathrm{min}} \{ \underset{\tau_v \in A(v)}{\mathrm{min}}  \{t(\tau_u, \tau_v, T)\}\}, \\
\end{equation}

\noindent where we select the minimal sleep delay value between sender $u$ and receiver $v$ after calculating the gaps between the active slots of the two nodes. 

We present the minimal sleep delay data aggregation scheduling scheme in Algorithm 1. As inputs, the communication graph $G(V,E)$ with sink node $s$, working period length $T$, the set of available channels $F$ and communication range $d$ are used. The output is all sensor nodes in the network are assigned time slots and allocated channels with collision-free. The aggregation tree construction phase is described from lines 1-8. Initially, sink node $s$ is added into the node set $V_T$, the set of transmission links $E_T$ is empty (lines 1-2). Based on the hop distance to the sink, the network is divided into $R+1$ layers, i.e., ${L_0,L_1,…,L_R}$. It means that only the sink is in layer $0$. 

\begin{algorithm}[!t]
    \caption{Data aggregation schedule}
    \label{algo: data_scheduling}
    \SetKwInOut{Input}{Input} 
    \SetKwInOut{Output}{Output}
    \Input{$G(V, E)$, $T$, sink node $s \in V, F, d$}
    \tcp{$T$: Working period length} 
    \tcp{$F$: Set of channels used for allocating the sensor nodes} 
    \tcp{$d$: Communication range of the sensors} 
    \Output{All sensor nodes in the network are assigned channels and time slots}
    \tcp{Aggregation tree construction phase}
    $V_T \leftarrow \{s\}$  \\
    $E_T \leftarrow \emptyset$ \\
    Divide the network into $R+1$ layers based on their hop distance to the sink $\{L_0, L_1,..., L_R\}$ \\
    \For {$i \in [1,R]$}
    {
        \For {$u\in L_i$}
        {
            $x \gets \underset{v\in N(u) \cap L_{i-1}}{\mathrm{argmin}} \{t_{min}(u,v,T)\}$ \\
            $p(u) \gets x$ \\
            $E_T \gets E_T \cup \{(u,x)\}$ \\
            $V_T \gets V_T \cup \{u\}$ \\
        }
        %\EndFor
    }
    \tcp{Data scheduling phase}
    Starting time $t_s \gets 0$ \\
    $V_{leaf} \gets \{u \in V_T \setminus \{s\} \mid C(u) = \emptyset\}$ \\
    \While {$V_{leaf} \neq \emptyset$}
    {
        Apply \textbf{DAS Algorithm [8]} to aggregate the data from all the leaf nodes to their parents in the tree; \\
        Remove scheduled nodes from the tree; \\
        Update $V_{leaf} \gets \{u \in V_T | C(u) = \emptyset\}$
    }
    %\EndWhile
    \Return All nodes in the networks are scheduled
\end{algorithm}

The algorithm then starts in top-down manner, from layer $L_1$ to layer $L_R$. At layer $L_i$, where $i \in [1,R]$ a node $u$ in the layer $u \in L_i$ is considered to check the minimal sleep delay with its neighbor nodes in the upper layer using Equation (1), i.e., $v \in L_{i-1} \cap N(u)$. The node that forms the smallest value of minimal sleep delay with $u$ is selected to be adopted as parent of $u$ (applying Equation (2)) and it is added into the aggregation tree (lines 5-8). So on and so forth, the sensor nodes in each layer are examined one by one. The process of the aggregation tree construction completes when all sensor nodes except the sink in the network adopt their parents.

In the data scheduling phase, scheduling time $t_s$ is initialized as $0$ (line 9). The algorithm obtains a set $V_{leaf}$  that includes all nodes having no children in the tree (line 10). In other words, these are called leaf nodes. This is an enhancement for candidate nodes selection compared to the data aggregation scheduling scheme in \cite{Jiao Xianlong} since that scheme only selects leaf nodes layer by layer at each scheduling round. After that, we apply the same procedure of data scheduling in \cite{Jiao Xianlong}. The set of leaf nodes $V_{leaf}$ is constantly updated at every scheduling round after valid nodes are scheduled. We achieve the scheduling when all nodes in the network except the sink are assigned channels and time slots (lines 11-14).

\begin{figure}[!t]
    \centering
    \begin{subfigure}{0.49\textwidth} 
        \includegraphics[scale=1]{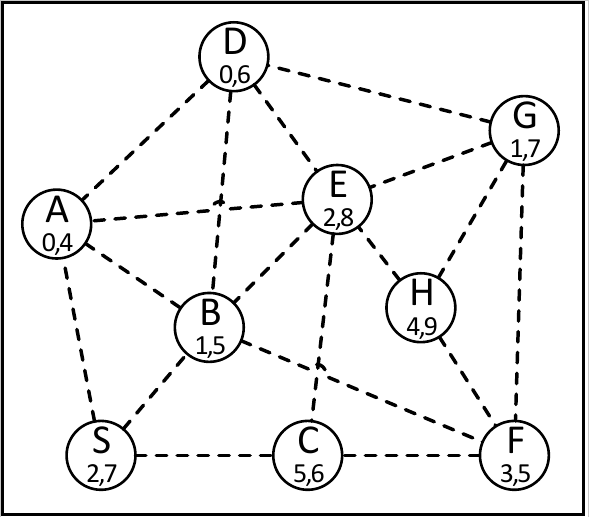}
        \caption{Communication graph} 
    \end{subfigure}  
    \hfill 
    \vspace{0.3cm}
    %\hspace{5mm}
    \begin{subfigure}{0.49\textwidth} 
        \includegraphics[scale=1]{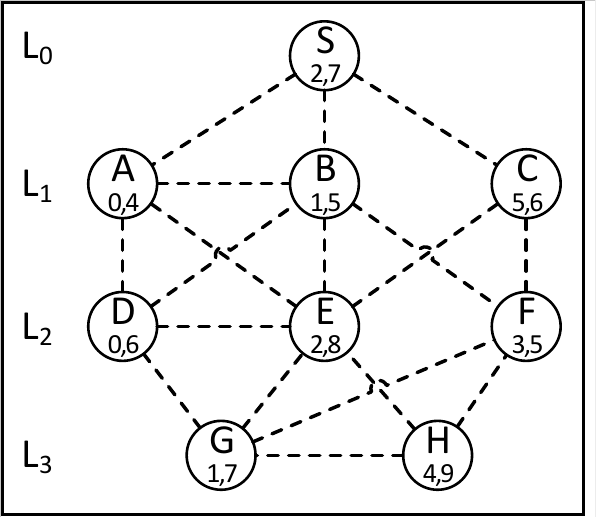}
        \caption{Network layering} 
    \end{subfigure}
    %\vspace{0.3cm}
    %\hspace{5mm}
    \begin{subfigure}{0.49\textwidth} 
        \includegraphics[scale=1]{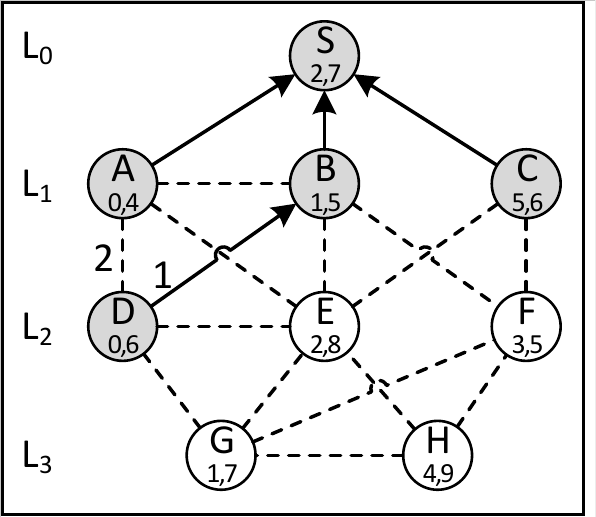}
        \caption{Aggregation tree construction process} 
    \end{subfigure}
    %\hspace{5mm}
    \begin{subfigure}{0.49\textwidth} 
        \includegraphics[scale=1]{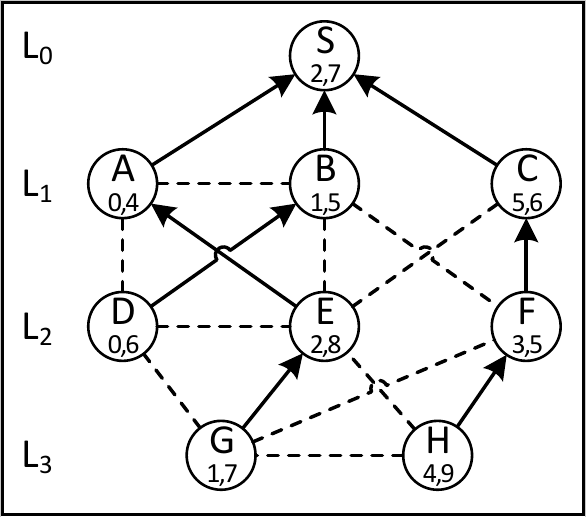}
        \caption{Complete aggregation tree} 
    \end{subfigure}
    \hspace{0.5cm}
    
    \begin{subfigure}{1\textwidth} 
        \centering
        \vspace{0.2cm}
        \includegraphics[scale=1]{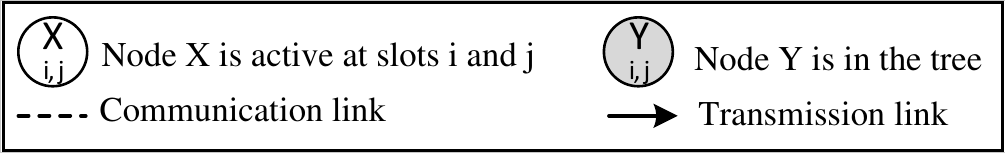}
    \end{subfigure}
    \caption{Example run for an aggregation tree construction process.}
    \label{fig: motivation example}
\end{figure}

An example of data aggregation tree construction is illustrated in Fig. 1. A sample communication graph consists of 9 sensor nodes ($S$ is the sink) is presented in Fig.1(a). Each sensor node is active at two slots at random and independently during a working period length $T= 10$. Then, the network is divided into layers based on the hop distance of sensors to the sink $S$ as shown in Fig. 1(b). The aggregation tree is constructed in a top-down manner. At layer $L_1$, nodes $A$, $B$, and $C$ have only one neighbor in layer $L_0$, so that they adopt the sink S as their parent. Then they are added into the tree. At layer $L_2$, the algorithm starts with node $D$. There are two neighbors in upper layer, i.e., nodes $A$ and $B$ in layer $L_1$, node $D$ takes $B$ as its parent since the minimal sleep delay between them is smallest (Fig. 1(c)). The algorithm works in same the way for the rest of the nodes in the network until all nodes adopt their parents, the completed tree is obtained as shown in Fig. 1(d).

\section{Experiment results}

In this section, we conduct extensive simulations to examine the effectiveness of our proposed schemes. We compare the performance of our proposed scheme, sleep Delay Data Aggregation Scheduling – DDAS, in terms of data aggregation latency, with an existing scheme \cite{Jiao Xianlong} that proposes a time-efficient data aggregation scheduling scheme, Novel Data Aggregation Scheduling scheme (NDAS), in multichannel duty-cycled sensor networks. We also compare the results to the SPT-DAS scheme, which builds the Shortest Path tree using Dijkstra's algorithm and employs the same scheduling scheme. To determine data aggregation latency, we generate 100 sensor networks randomly for each parameter variation. The data aggregation delay results in this section are an average of the 100 network topologies. In a working period length $T$ ranging from 10 to 70, sensor nodes are active in 1 to 7 slots. The sensor communication range is $d=20m$. In a $100 \times 100m^2$ area, the sensor nodes number can range between 50 and 800. The available channels range from 2 to 7.

\begin{figure}[!t]
    \centering
    \begin{subfigure}{0.495\textwidth} 
        \includegraphics[scale=0.25]{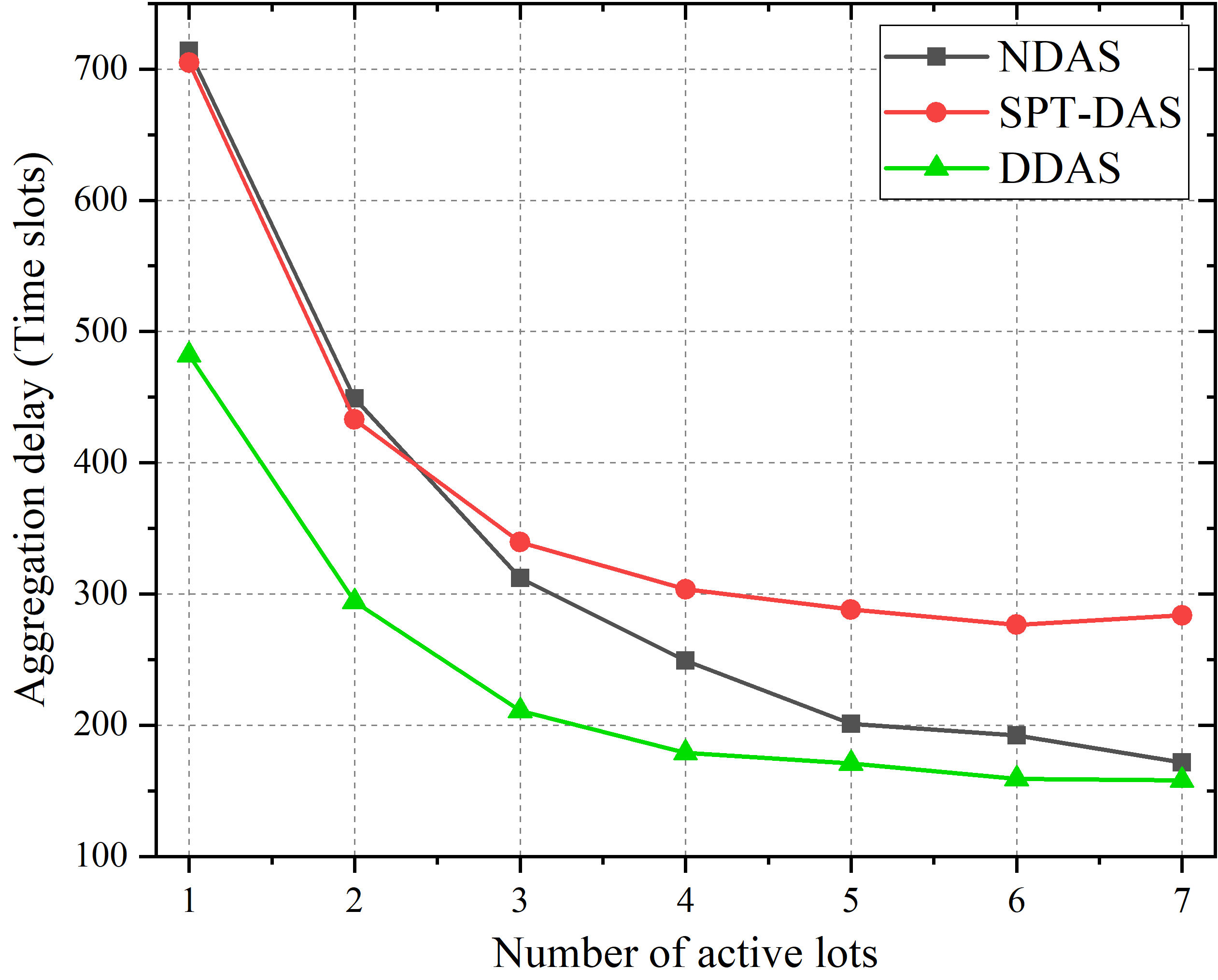}
        \caption{Number of active slots} 
    \end{subfigure}  
    \hfill 
    \begin{subfigure}{0.495\textwidth} 
        \includegraphics[scale=0.25]{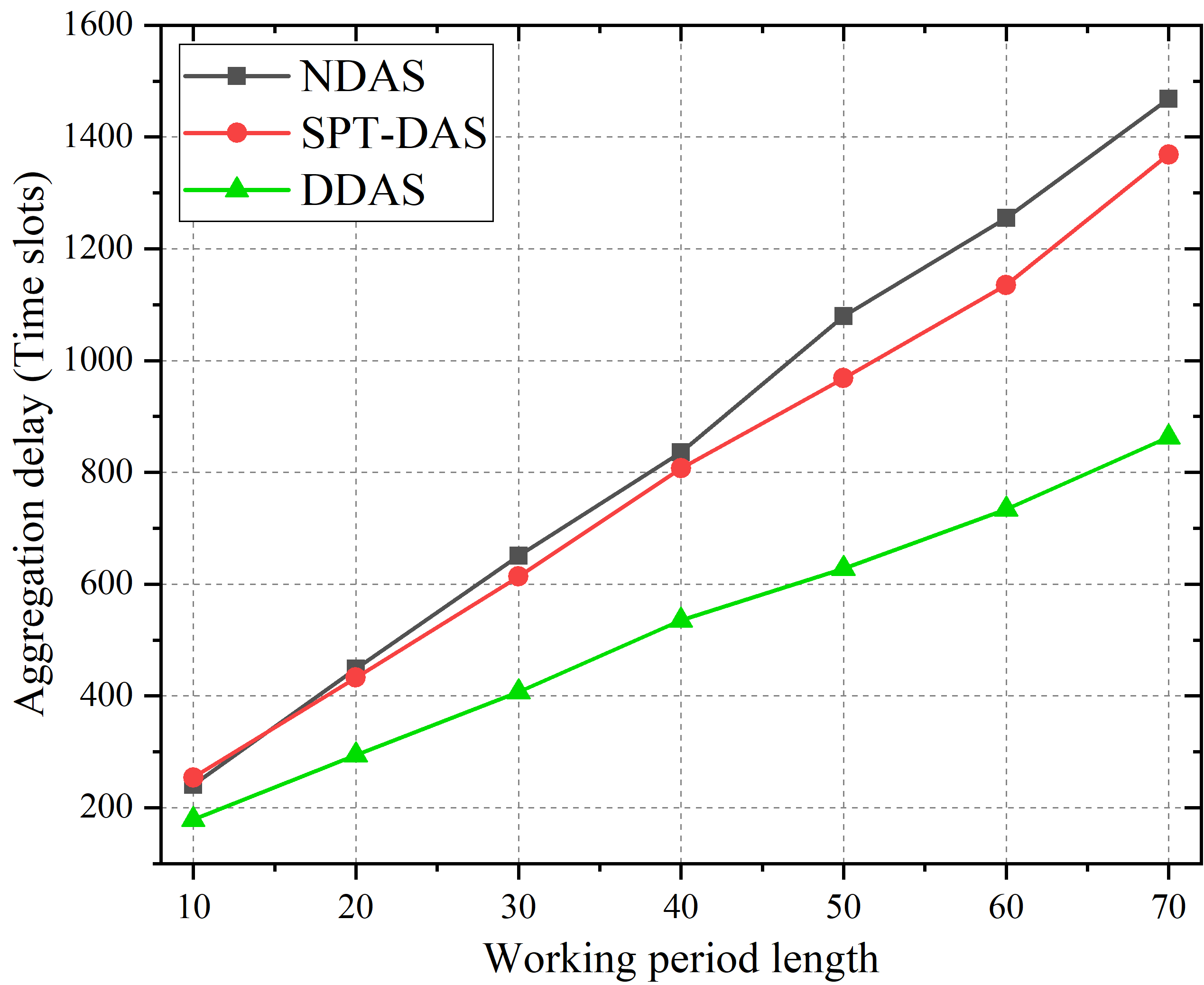}
        \caption{Working period length} 
    \end{subfigure}
    \caption{Aggregation delay affected by number of active slots and working period length.}
    \label{fig: motivation example}
\end{figure}

\textbf{Affect by active slots}. We run simulations to see how the number of active slots affects the data aggregation latency. We deploy 200 sensor node networks at random in the same size area. The number of active slots of each sensor node varies from 1 to 7 in a working period length $T = 20$. Figure 2(a) shows how data aggregation delay decreases as the number of active slots increases for all schemes. As the number of active slots at each sensor node increases, so does the number of simultaneous transmissions in a single working period. As a result, the overall aggregation delay is reduced. SPT-DAS appears to perform worse when compared to NDAS. Meanwhile, when the number of time slots is set to 2, the aggregation delay is reduced by up to 34\% when compared to DDAS.

\textbf{Affect by working period}. The number of nodes in each network is set to $N = 200$. Sensors operate in two slots over a working period length $T$ ranging from 10 to 70. The overall data aggregation delay of DDAS, SPT-DAS, and NDAS increases as the working period length increases, as shown in Fig. 2(b). Because nodes are designed to sleep for extended periods of time while waiting for their parents to become active in order to receive data from them. When we increase the working period length by the same number of active slots, the duty cycle is approximately 2.86-20\%. When compared to NDAS, DDAS achieves at most 41\% lower data aggregation latency at $T=50$, 60, and 70 because more nodes are scheduled at each scheduling iteration, resulting in more concurrent data transmissions and less data aggregation latency.

\begin{figure}[!t]
    \centering
    \begin{subfigure}{0.495\textwidth} 
        \includegraphics[scale=0.25]{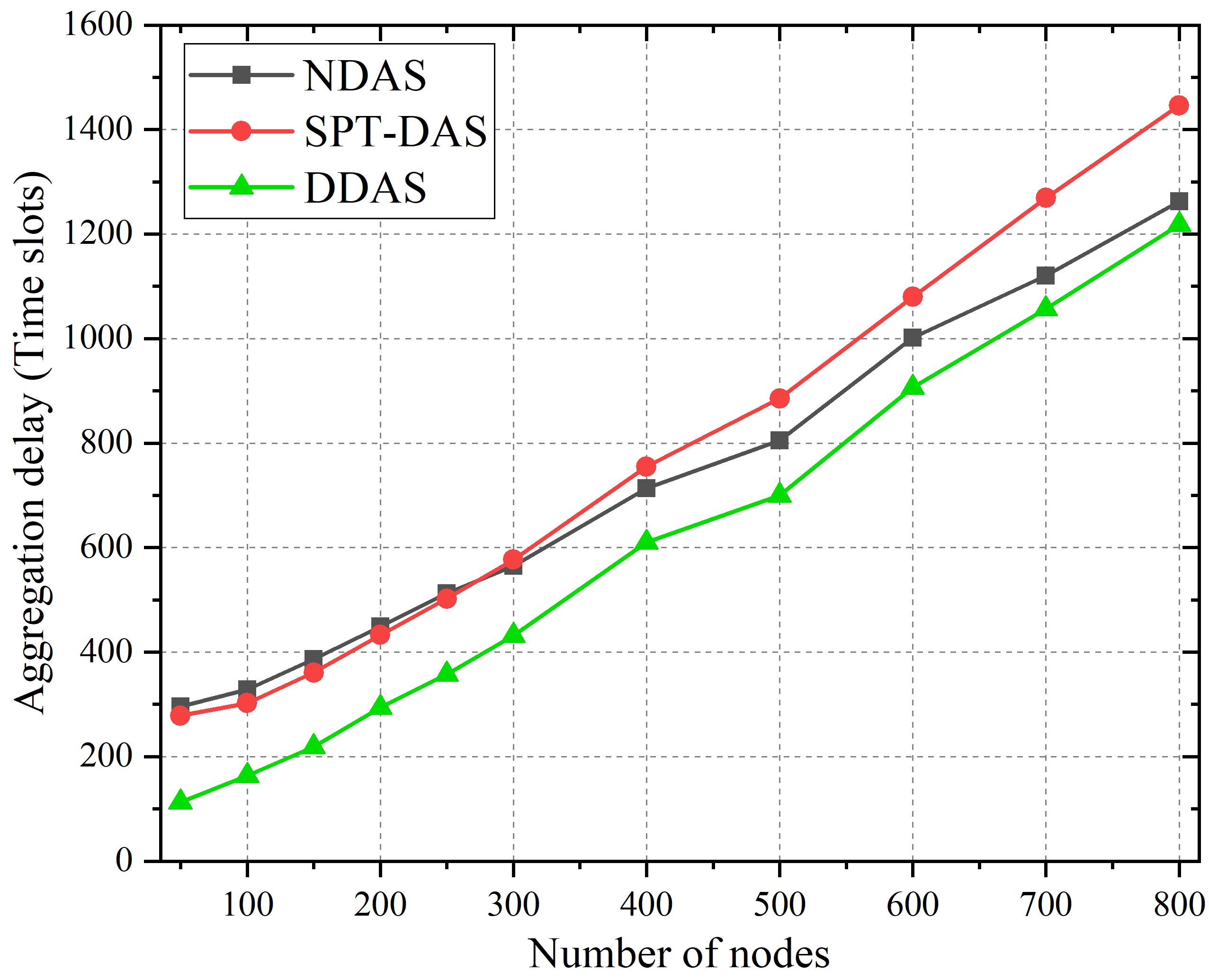}
        \caption{Number of nodes} 
    \end{subfigure}  
    \hfill 
    \begin{subfigure}{0.495\textwidth} 
        \includegraphics[scale=0.25]{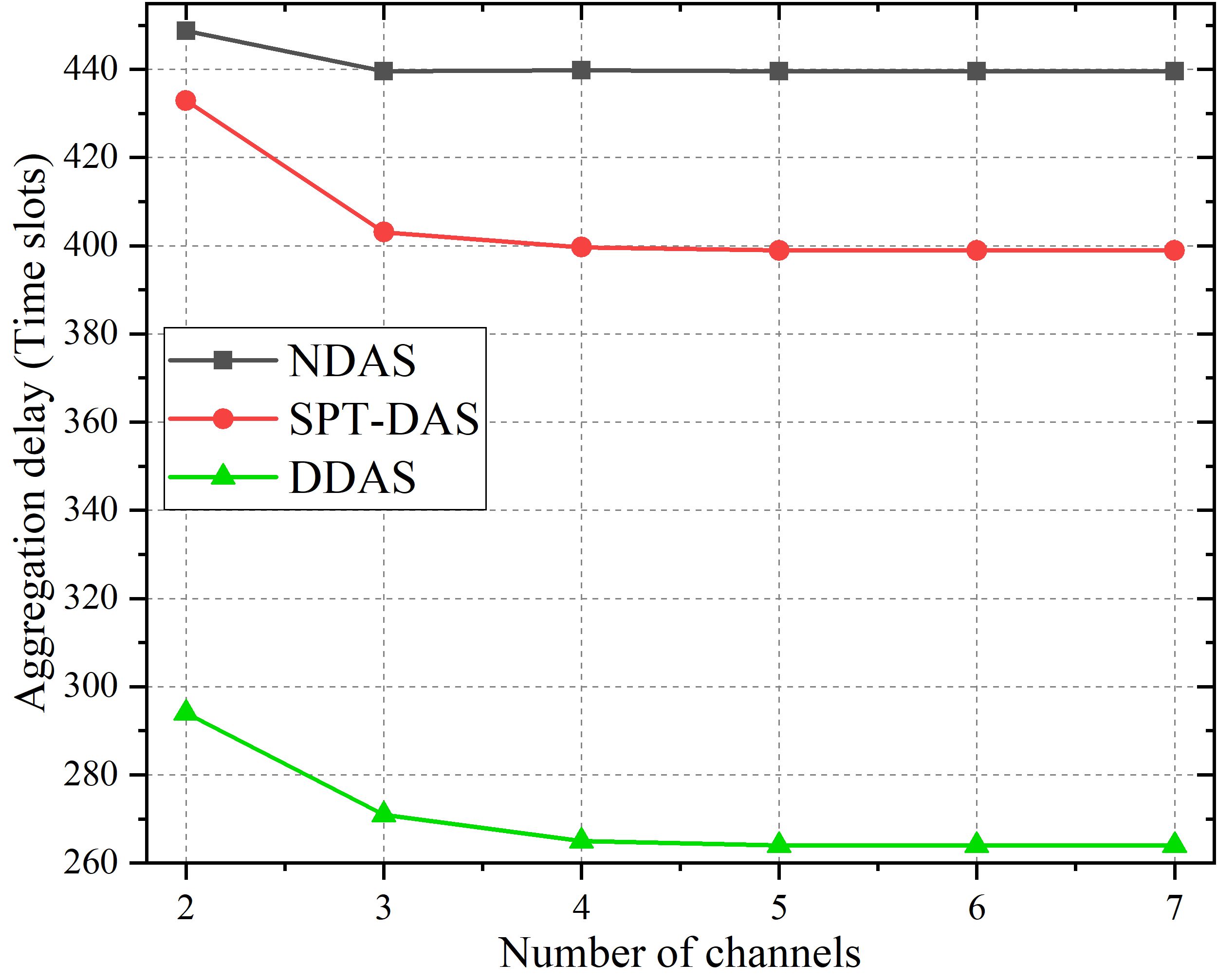}
        \caption{Number of channels} 
    \end{subfigure}
    \caption{Aggregation delay affected by number of sensor nodes and number of channels.}
    \label{fig: motivation example}
\end{figure}

\textbf{Affect by number of nodes}. The effect of node number is depicted in Fig. 3(a), where the number of nodes is increased from 50 to 800. In a working period length $T= 20$, sensor nodes are active at two slots. The three schemes' data aggregation delays increase as the number of nodes increases. Because as the number of nodes increases, more nodes must send the data to their parents. As a result, the data aggregation delay increases. At $N = 50$, DDAS lowers data aggregation latency by up to 62\% when compared with NDAS. This enhancement is also influenced by using an improved data scheduling and a different aggregation tree construction approaches.

\textbf{Affect by number of channels}. The usage of different channels further avoids interference while transmitting data between nodes. We deploy 200 sensor nodes the same area, they are active at 2 slots in $T= 20$. The number of channels is changed from 2 to 7. When using from 2 to 4 channels, the aggregation delay of all schemes reduce because more transmissions occurs simultaneously. However, with more available channels, the latency remains the same because the network is sparse, using more channels does not improve the performance. DDAS achieves better aggregation delay compared with NDAS and SPT-DAS schemes, i.e., at most 39.94\% and 33.83\%, respectively.

\section{Conclusion}
We propose the sleep Delay Data Aggregation Scheduling scheme (DDAS) in this paper, in which we use a new method of constructing the aggregation tree based on sleep delay among sensors, and then schedule all leaf nodes in the aggregation tree. The nodes that are scheduled, are then removed from the aggregation tree. We show an example run for the tree-building process. In addition, we conduct extensive experiments to evaluate the performance of DDAS to that of a modern scheme in terms of data aggregation delay. When compared to the start-of-the-art approach, our proposed scheme reduces data aggregation delay by up to 61\% depending on number of nodes, working period length, the number of active slots, and number of channels. In the future, we apply a break and joint approach to the current proposed scheme for further improving the data aggregation delay, which aids in maximizing parallel interference-free transmissions.

\vspace{1cm}

\noindent \textbf{Acknowledgement}. This work is partially supported by the Ministry of Science and ICT, BK21 FOUR, GITRC (IITP-2022-2015-0-00742), ICT Creative Consilience (IITP-2022-2020-0-01821), and Mid-career (NRF-2020R1A2C2008447).



\begin{thebibliography}{9}
%

\bibitem {Abbasian Dehkordi}
Abbasian Dehkordi, Soroush, Kamran Farajzadeh, Javad Rezazadeh, Reza Farahbakhsh, Kumbesan Sandrasegaran, and Masih Abbasian Dehkordi, ``A survey on data aggregation techniques in IoT sensor networks," Wireless Networks 26, no. 2 (2020): 1243-1263.

\bibitem {Wen Yean-Fu}
Wen, Yean-Fu, and Frank Yeong-Sung Lin, ``Energy-efficient data aggregation routing and duty-cycle scheduling in cluster-based sensor networks," In 2007 4th IEEE Consumer Communications and Networking Conference, pp. 95-99. IEEE, 2007.

\bibitem {KhadirKumar}
KhadirKumar, N., and A. Bharathi, ``Real time energy efficient data aggregation and scheduling scheme for WSN using ATL," Computer Communications 151 (2020): 202-207.

\bibitem {Chen Quan}
Chen, Quan, Hong Gao, Zhipeng Cai, Lianglun Cheng, and Jianzhong Li, ``Energy-collision aware data aggregation scheduling for energy harvesting sensor networks," In IEEE INFOCOM 2018-IEEE Conference on Computer Communications, pp. 117-125. IEEE, 2018.

\bibitem {Yu Bo}
Yu, Bo, and Jian-Zhong Li, ``Minimum-time aggregation scheduling in duty-cycled wireless sensor networks," Journal of Computer Science and Technology 26, no. 6 (2011): 962-970.

\bibitem {Chen Quan 2}
Chen, Quan, Hong Gao, Zhipeng Cai, Lianglun Cheng, and Jianzhong Li, ``Distributed low-latency data aggregation for duty-cycle wireless sensor networks," IEEE/ACM Transactions on Networking 26, no. 5 (2018): 2347-2360.

\bibitem {Jiao Xianlong}
Jiao, Xianlong, Wei Lou, Songtao Guo, Libin Yang, Xinxi Feng, Xiaodong Wang, and Guirong Chen, ``Delay efficient scheduling algorithms for data aggregation in multi-channel asynchronous duty-cycled WSNs," IEEE Transactions on Communications 67, no. 9 (2019): 6179-6192.

\bibitem{Shi Yi}
Shi, Yi, Y. Thomas Hou, Jia Liu, and Sastry Kompella, ``How to correctly use the protocol interference model for multi-hop wireless networks," In Proceedings of the tenth ACM international symposium on Mobile ad hoc networking and computing, pp. 239-248. 2009.

\bibitem{Chen Quan 3}
Chen, Quan, Hong Gao, Siyao Cheng, Jianzhong Li, and Zhipeng Cai, ``Distributed non-structure based data aggregation for duty-cycle wireless sensor networks," In IEEE INFOCOM 2017-IEEE Conference on Computer Communications, pp. 1-9. IEEE, 2017.

\end{thebibliography}
\end{document}